\documentclass[11pt,a4paper]{article}

\newcommand{\cpfversion}{Version 0.4.0}
\newcommand{\cpfbuild}{20170627}

\usepackage{latexsym}
\usepackage{amsmath}
\usepackage{amsfonts}
\usepackage{amsthm}
\usepackage{bbm}                                        % mathbbm is nicer than mathbb
\usepackage[utf8]{inputenc}
\usepackage[T1]{fontenc}
\usepackage{times}
\usepackage{listings}
\usepackage{microtype}
\usepackage{setspace}

\usepackage{color}
\definecolor{DarkBlue}{rgb}{0.0, 0.0, 0.5}
\definecolor{DarkRed}{rgb}{0.5, 0.0, 0.0}
\definecolor{DarkGreen}{rgb}{0.0, 0.5, 0.0}
\definecolor{DarkYellow}{rgb}{0.5, 0.5, 0.0}
\definecolor{Brown}{cmyk}{0.00,0.80,1.00,0.60}
\definecolor{DarkGreen}{cmyk}{0.64,0.00,0.95,0.60}
\definecolor{DarkBlue}{cmyk}{0.70,0.60,0.00,0.60}

\usepackage{colortbl}

\usepackage{cancel}
\usepackage{marvosym}

\definecolor{mediumgrey}{rgb}{0.8,0.8,0.8}
\definecolor{lightgrey}{rgb}{0.9,0.9,0.9}

\newif\iflayoutms
\layoutmsfalse

\newif\iflongversion
\longversionfalse

\newif\ifdevelopmentversion
\developmentversionfalse

\iflayoutms
\else

\usepackage{sectsty}
\allsectionsfont{\usefont{OT1}{phv}{}{n}\selectfont}            % Change font of section headers

\makeatletter
\newlength{\myFootnoteWidth}
\newlength{\myFootnoteLabel}
\renewcommand{\@makefntext}[1]{%
  \setlength{\myFootnoteWidth}{\columnwidth}%
  \addtolength{\myFootnoteWidth}{-\myFootnoteLabel}%
  \noindent\makebox[\myFootnoteLabel][r]{\@makefnmark\ }%
  \parbox[t]{\myFootnoteWidth}{#1}%
} \makeatother

\usepackage{fancyhdr}
\fi

\usepackage{longtable}

\usepackage[hyperindex = true, colorlinks = true, linktocpage = true, linkcolor=blue, citecolor=blue, bookmarks, pdftitle={Automatic Backward Differentiation for American Monte-Carlo Algorithms (Conditional Expectation)}, pdfauthor={Christian Fries}]{hyperref}

\usepackage{graphicx}
\usepackage{subfigure}

\iflayoutms
\else
\usepackage[aboveskip=10pt,belowskip=0pt,margin=10pt,labelfont=bf,format=plain]{caption}
\usepackage[nottoc]{tocbibind}
\fi

\newcounter{cpf_counter} 
\newcounter{cpfNumberOfFigures} 
\newcounter{cpfNumberOfTables} 

\newcommand{\indicatorfcn}{\mathrm{\mathbf{1}}}

\newcommand{\isbn}[2]{\href{http://www.amazon.com/dp/#1/finmath-20}{ISBN #2}}

\title{Automatic Backward Differentiation for American Monte-Carlo Algorithms (Conditional Expectation)}

\author{%
	Christian P.~Fries\\
	{\small \href{mailto:email@christian-fries.de}{email@christian-fries.de}}
}

\date{June 27, 2017}
\begin{document}

\pagestyle{fancy}               % of course

% Graphics
\DeclareGraphicsExtensions{.pdf,.jpg,.png}

%%%
%%% TITLE
%%%

\maketitle

\centerline{\small\href{http://www.christianfries.com/finmath/stoachasticautodiff}{\small\cpfversion}}

%%%
%%% ABSTRACT
%%%

\section*{Abstract}

In this note we derive the backward (automatic) differentiation (adjoint [automatic] differentiation) for an algorithm containing a conditional expectation operator. As an example we consider the backward algorithm as it is used in Bermudan product valuation, but the method is applicable in full generality.

The method relies on three simple properties:
\begin{enumerate}
	\item a forward or backward (automatic) differentiation of an algorithm containing a conditional expectation operator results in a linear combination of the conditional expectation operators;
	
	\item the differential of an expectation is the expectation of the differential $\frac{d}{dx} E(Y) = E(\frac{d}{dx}Y)$;
	
	\item if we are only interested in the expectation of the final result (as we are in all valuation problems), we may use $E(A \cdot E(B\vert\mathcal{F})) = E(E(A\vert\mathcal{F}) \cdot B)$, i.e., instead of applying the (conditional) expectation operator to a function of the underlying random variable (continuation values), it may be applied to the adjoint differential.
\end{enumerate}

The methodology not only allows for a very clean and simple implementation, but also offers the ability to use different conditional expectation estimators in the valuation and the differentiation.

%%%
%%% DISCLAIMER
%%%

\vfill

\begin{footnotesize}

\noindent \textbf{Disclaimer:} The views expressed in this work are the personal views of the authors and do not necessarily reflect the views or policies of current or previous employers. \newline
Feedback welcomed at \href{mailto:email@christian-fries.de}{email@christian-fries.de}.

\end{footnotesize}

%%%
%%% ACKNOWLEDGMENT
%%%

%\section*{Acknowledgment}

%We are grateful to Amine Chaieb for stimulating discussions.

\newpage

%%%
%%% TABLE OF CONTENTS
%%%

%\setcounter{tocdepth}{1}
\tableofcontents

%%%
%%% Here we go...
%%%

\newpage

\section{Introduction}
\label{sec:autodiffforamericanmontecarlo:introduction}

Given a filtered probability space $(\Omega, \mathbb{Q}, \{ \mathcal{F}_{t} \})$ we consider a Monte-Carlo simulation of a (time-discretised) stochastic process, i.e., we simulate sample path $\omega_{i}$ of sequences of random variables. Assuming that the drawings are uniform with respect to the measure $\mathbb{Q}$, this allows to approximate the unconditional expectation $\mathrm{E}^{\mathbb{Q}}(X)$ of a random variable $X$ via
\begin{equation*}
	\mathrm{E}^{\mathbb{Q}}(X) \ \approx \frac{1}{n} \sum_{l=0}^{n-1} X(\omega_{l}) \text{.}
\end{equation*}
However, the calculation of conditional expectations $\mathrm{E}^{\mathbb{Q}}( X \vert \mathcal{F}_{T_{i}})$ is known to be involved, since the Monte-Carlo simulation does not provide a discretization of the filtration $\{ \mathcal{F}_{t} \}$. For the calculation (estimation) of conditional expectations numerical approximations, like least-square regressions can be used. These methods are often referred to as \textit{American Monte-Carlo}, c.f.~\cite{FriesLectureNotes2007}.

Let $x$ denote a given model parameter used in the generation of the Monte-Carlo simulation. Given a valuation algorithm which calculates the unconditional expectation $\mathrm{E}^{\mathbb{Q}}(V)$ of a random variable $V$, where the calculation of $V$ involves one or more conditional expectations, we consider the calculation of the derivative $\frac{d}{d x} \mathrm{E}^{\mathbb{Q}}(V)$.

Using automatic differentiation (i.e., analytic differentiation and the chain rule), see~\cite{CapriottiGiles, GlassermanGilesAdjoint2006}, this would require a differentiation of the conditional expectation operator or - in practice - the approximation of the the conditional expectation operator. Hence, direct application of automatic differentiation would result in the differentiation of the approximation of the the conditional expectation operator, e.g., the regression, see~\cite{AntonovAADforBermudan2017, CapriottiJiangMarcinaAAD2016}.

First of all, this is know to be difficult due to the cross-sectional character of such approximations. In addition, it is in general questionable to differentiate an approximation, since it is not guaranteed that the differentiation of an approximation is a good approximation of the differential.

However, we can easily avoid this issue: First the differentiation of the conditional expectation can be replaced by a conditional expectation of a differentiation.
While this alone would present an improvement, it seems to make (backward) automatic differentiation much more complicated, since we would have to calculate a derivative for each argument of a conditional expectation. In the next section we show that we can simplify the algorithm by taking the conditional expectation of the adjoint.

%
% %%%%%%%%%%
%

\section{American Monte-Carlo and Bermudan Option Valuation}

We shortly given the definition of the backward algorithm. For details see~\cite{FriesLectureNotes2007}, we use a similar notation here.

Let $0 = T_{0} < T_{1} < \ldots < T_{n}$ denote a given time discretization. Let $V_{i}$, $i = 1, \ldots, n$ denote the time-$T_{i}$ numéraire relative values of given underlyings. Here $V_{i}$ are $\mathcal{F}_{T_{i}}$-measurable random variables. Then let $U_{i}$ be defined as follows:
\begin{align*}
	U_{n+1} & \ := \ 0 \\
	U_{i} & \ := B_{i}\left( \mathrm{E}^{\mathbb{Q}}\left( U_{i+1} \ \vert \mathcal{F}_{T_{i}} \right),  U_{i+1} , V_{i} \right) \text{,}
\end{align*}
where $B_{i}$ are arbitrary function.

\subsection{Bermudan Option Valuation}

For a Bermudan option $B_{i}$ is given by the optimal exercise criteria, i.e., $B_{i}(x,u,v) := G(x-v,u,v)$ with
\begin{equation*}
	G(y,u,v) \ := \ \left\{ \begin{array}{ll} u & \text{if y>0} \\ v & \text{else} \end{array} \right\} \text{.}
\end{equation*}

This defines a backward induction $i=n, n-1,\ldots, 1$ for $U_{i}$. For a Bermudan option we have that the \textit{unconditional} expectation (assuming that we have $N(T_{0}) = 1$ for the numéraire)
\begin{equation*}
	\mathrm{E}^{\mathbb{Q}}\left( U_{1} \right)
\end{equation*}
 is the (risk) neutral value of the Bermudan option with exercise dates $T_{1} < \ldots < T_{n}$ and exercise values $V_{i}$.

\subsection{Bermudan Digital Option Valuation}

For a Bermudan option the exercise is optimal. This allows for a popular ``trick'' to ignore the first order derivative of $G(y,u,v)$ with respect to $y$. Hence, for a Bermudan option it is not necessary to differentiate $B_{i}$ with respect to $x$.

However, this is just a special property of the function $G$ and only holds for first order derivatives. For example, the trick cannot applied to a Bermudan digital option, i.e., $B_{i}(x,u,v) := H(x-v,u,v)$ with
\begin{equation*}
	H(y,u,v) \ := \ \left\{ \begin{array}{ll} u & \text{if y>0} \\ 1 & \text{else} \end{array} \right\} \text{.}
\end{equation*}
We will use this product as a test case of the methodology in Section~\ref{sec:autodiffforamericanmontecarlo:numericalResults}.

\paragraph{Remarks:} The Bermudan \textit{digital} option shows a dependency on the first order derivative with respect to the  indicator condition $y$, since the expectation of the two outcomes $u$ and $1$ differs, i.e., we have a jump at $y=0$. For the classic Bermudan option we have that the expectation of $u$ and $v$ agree at $y=0$.

The (automatic) differentiation of a jump may appear as an issue, but it is possible to solve this elegantly in an stochastic automatic differentiation \cite{FriesChristianStochsticAutoDiff2017} and in this note we focus on the treatment of the conditional expectation (however, our numerical results demonstrate that the algorithm \cite{finmath-lib-automaticdifferentiation-extenstions} treats the discontinuity correctly, c.f.~\cite{FriesChristianStochsticAutoDiff2017}).

\section{Automatic Differentiation of the American-Monte-Carlo Backward Algorithm}

Let $z$ denote an arbitrary model parameter. We assume that the underlying values $V_{i}$ depend on $z$.

For $i=1, \ldots, n$ we have
\begin{align}
	\frac{\mathrm{d}}{\mathrm{d}z} U_{i}
	& \ = \ \frac{\mathrm{d}}{\mathrm{d}z} B_{i}\left( \mathrm{E}^{\mathbb{Q}}\left( U_{i+1} \ \vert \mathcal{F}_{T_{i}} \right),  U_{i+1} , V_{i} \right) \nonumber \\
	\intertext{and writing $B_{i} = B_{i}\left( \mathrm{E}^{\mathbb{Q}}\left( U_{i+1} \ \vert \mathcal{F}_{T_{i}} \right),  U_{i+1} , V_{i} \right)$ to avoid the lengthy argument list}
	\frac{\mathrm{d}}{\mathrm{d}z} U_{i}
	& \ = \ \frac{\mathrm{d} B_{i}}{\mathrm{d} x} \cdot \left( \frac{\mathrm{d}}{\mathrm{d}z} \mathrm{E}^{\mathbb{Q}}\left( U_{i+1} \ \vert \mathcal{F}_{T_{i}} \right) \right) + \frac{\mathrm{d} B_{i}}{\mathrm{d} u} \cdot \frac{\mathrm{d} U_{i+1}}{\mathrm{d}z} + \frac{\mathrm{d} B_{i}}{\mathrm{d} v} \cdot \frac{\mathrm{d} V_{i}}{\mathrm{d}z} \nonumber \\
	\intertext{and using $\frac{\mathrm{d}}{\mathrm{d}z} \mathrm{E}^{\mathbb{Q}}\left( U_{i} \right) \ = \ \mathrm{E}^{\mathbb{Q}}\left( \frac{\mathrm{d}}{\mathrm{d}z} U_{i} \right)$}
	\label{eq:autodiffforamericanmontecarlo:onestep}
	\frac{\mathrm{d}}{\mathrm{d}z} U_{i}
	& \ = \ \frac{\mathrm{d} B_{i}}{\mathrm{d} x} \cdot \mathrm{E}^{\mathbb{Q}}\left( \frac{\mathrm{d U_{i+1}}}{\mathrm{d}z} \ \vert \mathcal{F}_{T_{i}} \right) + \frac{\mathrm{d} B_{i}}{\mathrm{d} u} \cdot \frac{\mathrm{d} U_{i+1}}{\mathrm{d}z} + \frac{\mathrm{d} B_{i}}{\mathrm{d} v} \cdot \frac{\mathrm{d} V_{i}}{\mathrm{d}z}
\end{align}

\subsection{Forward Differentiation}

Applying this relation iteratively (plugging the expression of $\frac{\mathrm{d} U_{j+1}}{\mathrm{d}z}$ into the equation of $\frac{\mathrm{d} U_{i}}{\mathrm{d}z}$ for $j=i,\ldots,n-1$) gives
\begin{align*}
	\frac{\mathrm{d}}{\mathrm{d}z} U_{i}
	& \ = \ \sum_{j=i}^{n} \left( \left( \prod_{k=i}^{j-1} \frac{\mathrm{d} B_{k}}{\mathrm{d} u} \right) \frac{\mathrm{d} B_{j}}{\mathrm{d} x} \cdot \mathrm{E}^{\mathbb{Q}}\left( \frac{\mathrm{d} U_{j+1}}{\mathrm{d}z} \ \vert \mathcal{F}_{T_{j}} \right) + \left( \prod_{k=i}^{j-1} \frac{\mathrm{d} B_{k}}{\mathrm{d} u} \right) \frac{\mathrm{d} B_{j}}{\mathrm{d} v} \cdot \frac{\mathrm{d} V_{j}}{\mathrm{d}z} \right) \text{.}
\end{align*}
To shorten notation, let
\begin{equation*}
	A_{i,j} = \left( \prod_{k=i}^{j-1} \frac{\mathrm{d} B_{k}}{\mathrm{d} u} \right) \frac{\mathrm{d} B_{j}}{\mathrm{d} x} \qquad C_{i,j} = \left( \prod_{k=i}^{j-1} \frac{\mathrm{d} B_{k}}{\mathrm{d} u} \right) \frac{\mathrm{d} B_{j}}{\mathrm{d} v}
\end{equation*}
such that
\begin{equation}
	\label{eq:autodiffforamericanmontecarlo:forward}
	\frac{\mathrm{d}U_{i}}{\mathrm{d}z} \ = \ \sum_{j=i}^{n} \left( A_{i,j} \cdot \mathrm{E}^{\mathbb{Q}}\left( \frac{\mathrm{d} U_{j+1}}{\mathrm{d}z} \ \vert \mathcal{F}_{T_{j}} \right) +  C_{i,j} \cdot \frac{\mathrm{d} V_{j}}{\mathrm{d}z} \right) \text{.}
\end{equation}
This last equation would be natural in a forward (automatic) differentiation, since we calculate $0 = \frac{\mathrm{d}U_{n+1}}{\mathrm{d}z}$, $\frac{\mathrm{d}U_{n}}{\mathrm{d}z}$, $\frac{\mathrm{d}U_{n-1}}{\mathrm{d}z}$, \ldots, together with $\frac{\mathrm{d} V_{j}}{\mathrm{d}z}$ forwardly. Note that forward here refers to the order of the operations in the algorithms (which runs backward over the indices $j$).

\subsection{Backward Differentiation}

For the application of a backward (adjoint) automatic differentiation Equation~\eqref{eq:autodiffforamericanmontecarlo:forward} would require a mixture of backward differentiation for  that we first calculate $\frac{\mathrm{d}U_{j}}{\mathrm{d}z}$ and followed by a forward application of~\eqref{eq:autodiffforamericanmontecarlo:forward}. However, we can calculate the derivative in a single backward differentiation sweep:

We start with \eqref{eq:autodiffforamericanmontecarlo:onestep} for $i=1$. Since we are only interest in $\frac{\mathrm{d}}{\mathrm{d}z} \mathrm{E}^{\mathbb{Q}}\left( U_{1} \right) = \mathrm{E}^{\mathbb{Q}}\left( \frac{\mathrm{d}}{\mathrm{d}z} U_{1} \right)$, we can take expectation and get
\begin{align*}
	\mathrm{E}^{\mathbb{Q}}\left( \frac{\mathrm{d}}{\mathrm{d}z} U_{1} \right)
	& \ = \ \mathrm{E}^{\mathbb{Q}}\left( \frac{\mathrm{d} B_{1}}{\mathrm{d} x} \cdot \mathrm{E}^{\mathbb{Q}}\left( \frac{\mathrm{d U_{2}}}{\mathrm{d}z} \ \vert \mathcal{F}_{T_{i}} \right) + \frac{\mathrm{d} B_{1}}{\mathrm{d} u} \cdot \frac{\mathrm{d} U_{2}}{\mathrm{d}z} + \frac{\mathrm{d} B_{1}}{\mathrm{d} v} \cdot \frac{\mathrm{d} V_{1}}{\mathrm{d}z} \right)
\end{align*}
Now we may use
\begin{equation*}
	\mathrm{E}^{\mathbb{Q}}\left( \frac{\mathrm{d} B_{1}}{\mathrm{d} x} \cdot \mathrm{E}^{\mathbb{Q}}\left( \frac{\mathrm{d U_{2}}}{\mathrm{d}z} \ \vert \mathcal{F}_{T_{1}} \right) \right)
	\ = \ \mathrm{E}^{\mathbb{Q}}\left( \mathrm{E}^{\mathbb{Q}}\left( \frac{\mathrm{d} B_{1}}{\mathrm{d} x} \ \vert \mathcal{F}_{T_{1}} \right) \cdot \frac{\mathrm{d} U_{2}}{\mathrm{d}z} \right)
\end{equation*}
to get
\begin{equation}
	\label{eq:autodiffforamericanmontecarlo:forwardtwostep}
	\mathrm{E}^{\mathbb{Q}}\left( \frac{\mathrm{d}}{\mathrm{d}z} U_{1} \right)
	\ = \ \mathrm{E}^{\mathbb{Q}}\left( \left(\mathrm{E}^{\mathbb{Q}}\left( \frac{\mathrm{d} B_{1}}{\mathrm{d} x} \ \vert \mathcal{F}_{T_{1}} \right) + \frac{\mathrm{d} B_{1}}{\mathrm{d} u}\right) \cdot \frac{\mathrm{d} U_{2}}{\mathrm{d}z} + \frac{\mathrm{d} B_{1}}{\mathrm{d} v} \cdot \frac{\mathrm{d} V_{1}}{\mathrm{d}z} \right) \text{.}
\end{equation}
Plugging \eqref{eq:autodiffforamericanmontecarlo:onestep} into \eqref{eq:autodiffforamericanmontecarlo:forwardtwostep} and repeating the previous argument for $i=2,\ldots, k-1$ we get iteratively the forward equation
\begin{align*}
	\mathrm{E}^{\mathbb{Q}}\left( \frac{\mathrm{d}}{\mathrm{d}z} U_{1} \right)
	& \ = \ \mathrm{E}^{\mathbb{Q}}\left( A^{*}_{1,k} \cdot \frac{\mathrm{d} U_{k+1}}{\mathrm{d}z} + \sum_{j=1}^{k} C^{*}_{1,j} \cdot \frac{\mathrm{d} V_{j}}{\mathrm{d}z} \right) \text{.}
\end{align*}
where
\begin{align*}
	A^{*}_{1,i} & \ = \ \mathrm{E}^{\mathbb{Q}}\left( A^{*}_{1,i-1} \frac{\mathrm{d} B_{i}}{\mathrm{d} x} \ \vert \mathcal{F}_{T_{i}} \right) + A^{*}_{1,i-1} \frac{\mathrm{d} B_{i}}{\mathrm{d} u} \\
	C^{*}_{1,i} & \ = \ A^{*}_{1,i-1} \frac{\mathrm{d} B_{i}}{\mathrm{d} v} \\
	A^{*}_{1,0} & \ = \ 1
\end{align*}
Using $i=n$ we have with $U_{n+1} = 0$ that
\begin{align*}
	\mathrm{E}^{\mathbb{Q}}\left( \frac{\mathrm{d}}{\mathrm{d}z} U_{1} \right)
	& \ = \ \mathrm{E}^{\mathbb{Q}}\left( \sum_{j=1}^{n} C^{*}_{1,j} \cdot \frac{\mathrm{d} V_{j}}{\mathrm{d}z} \right) \text{.}
\end{align*}

The recursive definitions of $A^{*}_{1,i}$, $C^{*}_{1,i}$ have an intuitive interpretation in a backward (automatic) differentiation algorithm: in the algorithm the conditional expectation operator on $U_{i+1}$ is replaced by taking the conditional expectation of the adjoint differential.

\newpage

\section{Numerical Results}
\label{sec:autodiffforamericanmontecarlo:numericalResults}

As a test case we calculate the delta of a Bermudan digital option. This product pays
\begin{equation*}
	1	\qquad \text{\ if\ } S(T_{i})-K_{i} > \tilde{U}(T_{i}) \text{\ in $T_{i}$ if no payout has been occurred,}
\end{equation*}
where $\tilde{U}(T)$ is the time $T$ value of the future payoffs, for $T_{1},\ldots,T_{n}$. Note that $\tilde{U}(T_{n}) = 0$, such that the last payment is a digital option.

This product is an ideal test-case: the valuation of $\tilde{U}(T_{i})$ is a conditional expectation. In addition conditional expectation only appears in the indicator function, such that keeping the exercise boundary (the condition) fixed, would result in a delta of $0$. On the other hand, the delta of a digital option payoff is only driven by the movement of the indicator function, since
\begin{equation*}
	\frac{\mathrm{d}}{\mathrm{d} S_{0}} \mathrm{E}( \indicatorfcn(f(S(T)) > 0) ) = \phi(f^{-1}(0)) \frac{\mathrm{d f(S)}}{\mathrm{d} S_{0}} \text{,}
\end{equation*}
where $\phi$ is the probability density of $S$ and $\indicatorfcn(\cdot)$ the indicator function. See~\cite{FriesLectureNotes2007}

\begin{figure}[hbtp]
	\begin{center}
		\includegraphics[scale=0.70]{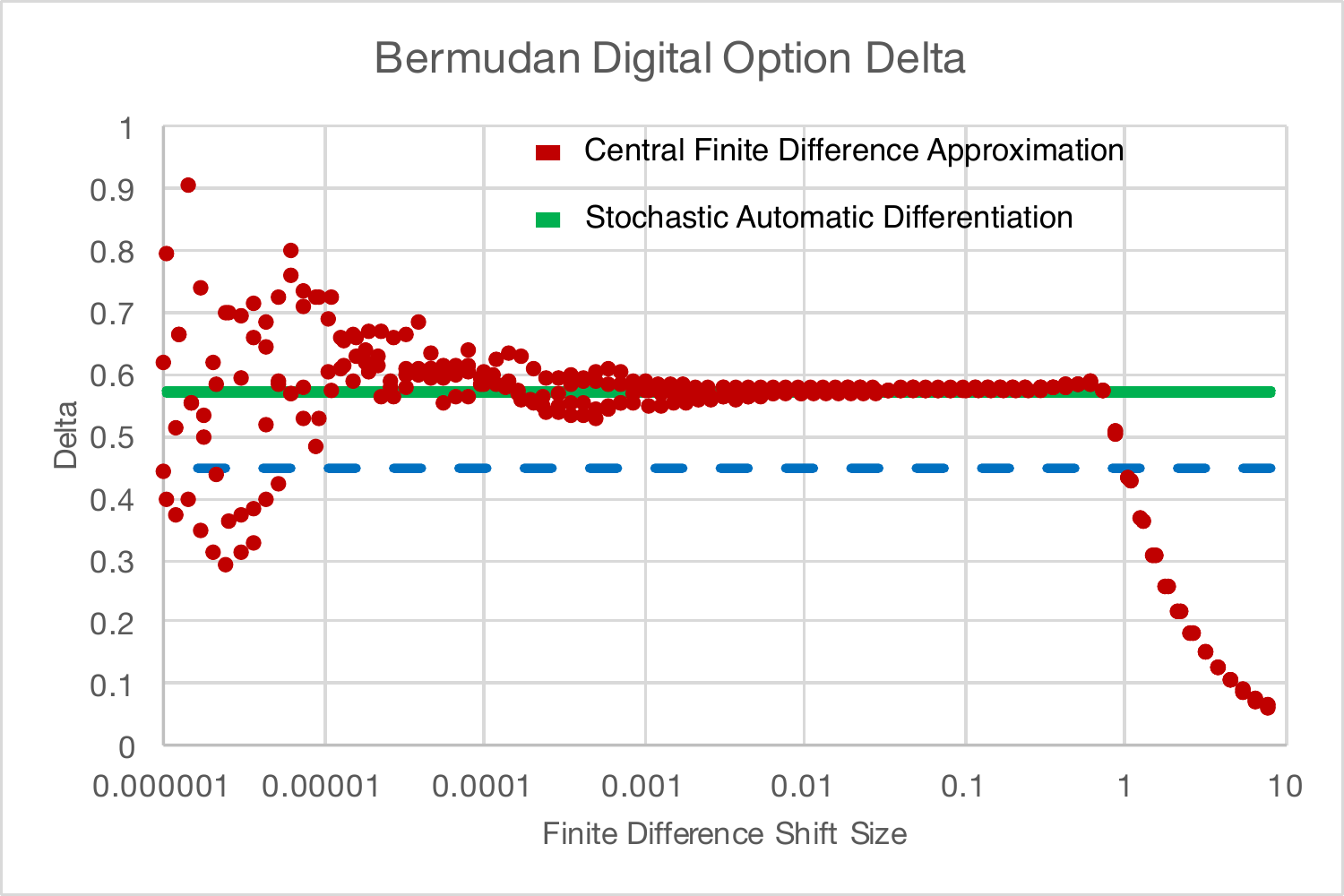}
		\caption[
		]{
			Delta of a Bermudan digital option using finite differences (red) and stochastic AAD. The calculations were repeated with 4 different Monte-Carlo random number seeds (hence there 4 red data series and 4 data series).
			The finite difference approximation is biased for large shifts size
			The blue dashed line depicts the value of the delta of the conditional expectation operator is ignored. This generates a bias du to the correlation of derivative and underlying.
		}
		\label{fig:correctingOptionPriceForConvexitiy}
	\end{center}
	\addtocounter{cpfNumberOfFigures}{1}
\end{figure}

The results are depicted in Figure~\ref{fig:correctingOptionPriceForConvexitiy} for $T_{1} = 1$, $T_{2} = 2$, $T_{3} = 3$, $T_{4} = 4$, and $K_{1} = 0.5$, $K_{2} = 0.6$, $K_{3} = 0.8$, $K_{4} = 1.0$ for a model with $S_{0} = 1$.

The implementation of this test case is available in~\cite{finmath-lib-automaticdifferentiation-extenstions}.

\section{Implementation Design}

Using the framework described in~\cite{FriesChristianStochsticAutoDiff2017} we are able to implement the methodology with a minimum of code complexity: we require only two additional lines of code:
In~\cite{finmath-lib} all random variable objects of a Monte-Carlo simulation are implementing a common interface \texttt{RandomVariableInterface}. This interface offers methods for arithmetic operators like \texttt{add}, \texttt{sub}, \texttt{mult}, \texttt{exp}, but also an operator
\begin{lstlisting}[frame=lr,texcl=true]
	RandomVariableInterface getConditionalExpectation(ConditionalExpectationEstimatorInterface estimator);
\end{lstlisting}
The default implementation of this method is just
\begin{lstlisting}[frame=lr,texcl=true]
	default RandomVariableInterface getConditionalExpectation(ConditionalExpectationEstimatorInterface estimator)
	{
		return estimator.getConditionalExpectation(this);
	}
\end{lstlisting}
In other words: all conditional expectations of random variables have to be called through this method - which enables us to track conditional expectation in the operator tree.

To enable adjoint automatic differentiation we inject a special implementation of \texttt{RandomVariableInterface} which records the operator tree and can perform the AAD. The implementation of the differential of the operator \texttt{getConditionalExpectation} now consists of two parts: First, the partial derivative of the operator with respect to its arguments is set to 1.0:
\begin{lstlisting}[frame=lr,texcl=true]
	case CONDITIONAL_EXPECTATION:
		return new RandomVariable(1.0);
\end{lstlisting}
- this implies that for the arguments the conditional expectation is replaced by the identity operator.
Second, the backward propagation of the differential contains the additional check for the operator, applying it to the differential if required:
\begin{lstlisting}[frame=lr,texcl=true]
	if(operatorType == OperatorType.CONDITIONAL_EXPECTATION) {
		ConditionalExpectationEstimatorInterface estimator = (ConditionalExpectationEstimatorInterface)operator;
		derivative = estimator.getConditionalExpectation(derivative);
	}
\end{lstlisting}

\section{Conclusion}

In this note we presented a simple modification of the backward automatic differentiation of algorithms which require the calculation (estimation) of a conditional expectation: the conditional expectation operator has to be applied to the adjoint differential, while the remaining algorithm remains unchanged.

Our method has important advantages over a direct differentiation of the conditional expectation estimator (\cite{AntonovAADforBermudan2017, CapriottiJiangMarcinaAAD2016}):
\begin{enumerate}
	\item Do not differentiate an approximation: It is often not clear that differentiation of an approximation given a good approximation of the differential. In our approach it is not necessary to differentiate the approximation of the conditional expectation operator. A simple example is if the conditional expectations is approximated by fixed bins, which results in a pice-wise constant approximation (see~\cite{FriesLectureNotes2007}) and the (automatic) differentiation of the conditional expectation would be zero.
	
	\item The approach allows to use a different approximations of the conditional expectation operator in the valuation and in the algorithmic differentiation. This can improve the accuracy of the differentiation. For example, it is possible to choose different regression basis functions.
\end{enumerate}

Most importantly, the approach greatly simplifies the implementation of the algorithm. In \cite{finmath-lib-automaticdifferentiation-extenstions} the implementation basically consists of adding two lines of code.

\newpage
%\bibliography{papers.bib}
%\bibliographystyle{model2-names}
%\bibliographystyle{elsarticle-harv}

%\newpage

\section*{Notes}
%\addcontentsline{toc}{section}{Notes}

\subsection*{Suggested Citation}

\begin{itemize}
	\item[] \sloppypar \textsc{Fries, Christian P.}: Automatic Differentiation of the American-Monte-Carlo Backward Algorithm. (June, 2017). \url{https://ssrn.com/abstract=3000822}
	\newline
	\url{http://www.christian-fries.de/finmath/stoachasticautodiff}
\end{itemize}

\subsection*{Classification}

{\small

\noindent Classification:
\href{http://www.ams.org/msc/}{MSC-class}: 65C05 (Primary)%, 68U20, 60H35 (Secondary).
\\
\phantom{Classification:}
\href{http://www.acm.org/class/1998/ccs98.html}{ACM-class}: G.1.4; G.3; I.6.8.\\
\phantom{Classification:}
\href{http://www.aeaweb.org/journal/jel_class_system.html}{JEL-class}: C15, G13.\\

%\noindent Keywords:
%Monte-Carlo Simulation,
%Valuation,
%Stress Test,
%Variance Reduction,
%\\
%\phantom{Keywords:\ }
%Boundary Conditions,
%Numerical Schemes,
%CEV,
%Variance Gamma
%}
%% MSC 65C05 Monte-Carlo methods
%% JEL C15

%%\newpage

%\

%\vfill

%\bigskip
%\centerline{\small\thepage \ pages. \thecpfNumberOfFigures \ figures. \thecpfNumberOfTables \ tables.}

\end{document}